**Title: The price elasticity of Gleevec in patients with Chronic Myeloid Leukemia enrolled in Medicare Part D: Evidence from a regression discontinuity design**

**Running Title: Price elasticity of Gleevec in leukemia patients on Part D**

Samantha E. Clark[1], Ruth Etzioni[2], Jerry Radich[2], Zachary Marcum[3], Anirban Basu[1]

1. The Comparative Health Outcomes, Policy, and Economics (CHOICE) Institute, University of Washington, Seattle, WA
2. Fred Hutchinson Cancer Research Center, Seattle, Washington; University of Washington School of Medicine, Seattle, WA
3. Aetion Inc., Boston, MA

**Funding:** This research was supported by a research grant from the National Institutes of Health (NIH). The funders had no role in study design, data collection and analysis, decision to publish, or preparation of the manuscript.

**Word count: 3,987**

# ABSTRACT

**Objective**

To assess the price elasticity of branded imatinib in chronic myeloid leukemia (CML) patients on Medicare Part D to determine if high out-of-pocket payments (OOP) are driving the substantial levels of non-adherence observed in this population.

**Data sources and study setting**

We use data from the TriNetX Diamond Network (TDN) United States database for the period from first availability in 2011 through the end of patent exclusivity following the introduction of generic imatinib in early 2016.

**Study design**

We implement a fuzzy regression discontinuity design to separately estimate the effect of Medicare Part D enrollment at age 65 on adherence and OOP in newly-diagnosed CML patients initiating branded imatinib. The corresponding price elasticity of demand (PED) is estimated and results are assessed across a variety of specifications and robustness checks.

**Data collection/extraction methods**

Data from eligible patients following the application of inclusion and exclusion criteria were analyzed.

**Principal findings**

Our analysis suggests that there is a significant increase in initial OOP of $232 (95% Confidence interval (CI): $102 to $362) for individuals that enrolled in Part D due to expanded eligibility at age 65. The relatively smaller and non-significant decrease in adherence of only 6 percentage points (95% CI: -0.21 to 0.08) led to a PED of -0.02 (95% CI: -0.056, 0.015).

**Conclusion**

This study provides evidence regarding the financial impact of coinsurance-based benefit designs on Medicare-age patients with CML initiating branded imatinib. Results indicate that factors besides high OOP are driving the substantial non-adherence observed in this population and add to the growing literature on PED for specialty drugs.

> What is known on this topic
>
> - The unique characteristics of specialty drugs means that patient adherence may be more sensitive with respect to out-of-pocket payments (OOP) than what has been observed for non-specialty drugs.
> - Imatinib is a highly effective and expensive oral specialty drug that is underutilized by chronic myeloid leukemia (CML) patients, most of whom are Medicare-age.
> - It's suspected that the high OOP for branded imatinib in coinsurance-based systems like Medicare Part D is a primary driver of the 20-30% nonadherence rate observed in this population.
>
> What this study adds
>
> - Despite a large increase in OOP associated with enrollment in Part D, adherence in CML patients was relatively unaffected.
> - Coinsurance-based benefit plans may be putting patients that are on effective, but expensive, specialty drugs like branded imatinib at unnecessary financial risk.
> - Future work should consider factors besides high OOP as drivers of the substantial non-adherence to imatinib observed in the CML population.

**INTRODUCTION**

The use of specialty medications has grown dramatically in the United States over the last 20 years (1). While many specialty drugs confer survival and quality of life improvements, their high prices have drawn increasing scrutiny from media, policymakers, and patients. This is especially true within the Medicare Part D program, where brand-name specialty drugs accounted for only 1 percent of all prescriptions dispensed but 30 percent of total net spending on prescription drugs in 2015 (2). Due to the benefit structure in Part D, where cost-sharing is based on list price, patients prescribed specialty drugs often enter the catastrophic coverage phase ($7,050 TrOOP threshold in 2022) after their first prescription fill (3). With 5% coinsurance and no spending cap in this phase, many beneficiaries face high annual medication costs (4). The financial burden imposed by these treatments can have important behavioral consequences, including failure to initiate, medication rationing, and discontinuation (5,6).

As the increasing strain of high OOP in Medicare beneficiaries has become impossible to ignore, there have been numerous calls for a redesign of the Part D benefit structure (4). The recently implemented Inflation Reduction Act (IRA) represents an important step towards addressing many of these concerns; most notably, by eliminating 5% coinsurance in catastrophic coverage and capping total OOPs at $2,000 in 2024 and 2025, respectively (9). Rebuffing the types of legal challenges and threats to repeal that the Affordable Care Act has and continues to face will require rigorous evidence on the potential impact of reverting to a coinsurance-based payment structure with no spending cap. Additionally, as private insurers move towards benefit plan designs with higher deductibles and coinsurance as a method of cost containment, it is important that the full implications of shifting costs onto patients are known (3,10–12).

Key to future policy discussions will be a better understanding of how high OOP affects medication adherence in patients. Analyses since the seminal RAND Health Insurance Experiment study support initial findings regarding the inelasticity of demand for prescription drugs; however, their focus was largely on non-specialty drugs for which the corresponding financial burden is low (13–15).

Evidence from the literature assessing specialty cancer therapies is more mixed, with some analyses suggesting a higher elasticity for these drugs than for their low-cost counterparts (16). In a 2016 review on cost sharing and specialty drug use, Doshi et al found that high OOP was more strongly associated with non-initiation and abandonment than adherence, which appeared to be insensitive to price (5). These findings are supported by similar studies indicating that patient demand for specialty drugs is fairly inelastic, with estimates ranging from -0.04 to -0.26 (6,15,16). Recent work however, demonstrated that patients could be more responsive to price than initially thought. Using an instrumental variable design, Jung et al identified the overall elasticity for specialty drugs to be between -0.72 and -0.75, with leukemia-specific estimates of -0.96 to -0.99 (17).

The use of imatinib, a tyrosine kinase inhibitor (TKI), in the treatment of chronic myeloid leukemia (CML) provides an ideal case study for assessing the relationship between list price, OOP, and adherence in

Medicare Part D since almost half of patients are diagnosed over the age of 65 (18). In addition to being one of the most successful oral chemotherapy agents ever developed, branded imatinib is also among the costliest (10,19). Although OOP has been identified as a potential driver of non-adherence in the approximately 20-30% of CML patients who exhibit this behavior, a definitive relationship between the two has not been established (20–22, 27).

This work contributes to the growing literature on the price elasticity of specialty drugs by using a regression discontinuity design to estimate the effect of the Medicare Part D benefit structure on initial OOP and adherence in patients with CML diagnosed right before and right after age 65. Our analytic approach combined with the information available in our data allow us to account for previously unaddressed endogeneity associated with plan choice (patients with higher drug costs tend to opt for more generous plans) and the distortionary effect of coupon use (5,17,26–28).

The primary aim of this study is to assess the price elasticity of branded imatinib in chronic myeloid leukemia (CML) patients on Medicare Part D to determine if high out-of-pocket payments (OOP) are driving the substantial levels of non-adherence observed in this population. Results provide richer context to previous associational studies that have identified a link between OOP and adherence in this population and add to the growing literature on the price elasticity of demand for specialty drugs.

**MATERIALS AND METHODS**

**Data**

We use data from the TriNetX Diamond Network (TDN) database for the period from first availability in 2011 through the end of patent exclusivity following the introduction of generic imatinib in early 2016. TDN comprises a convenience sample of claims and encounter data from 1.8 million providers representing 99 percent of U.S. health plans, with longitudinal data covering over 200 million patients (30). Data are pulled from open claims clearinghouses and include de-identified medical and pharmacy claims for Commercial, Disability/Workers Compensation, Dual Coverage, Medicare, Medicaid, and VA/Other plans.

The decision to use TDN for the primary analysis rather than other available claims databases (SEER-Medicare, MarketScan) is based on the following considerations: ability to capture hypothesized discontinuities across comparable pre- and post-Medicare populations and inclusion of coupon use as a form of payment assistance.

**Study design**

The study period consists of the time from the first branded imatinib fill, defined as the index date, to the end of three-months of follow-up, with the six months preceding the index date serving as a washout period for confirming incident diagnoses (no previous claims with CML diagnosis or treatment) and the baseline period for covariate measurement. A new user design is adopted to account for survivor bias (the most elastic patients will have the worst outcomes), healthy user bias (non-adherers are more likely to switch treatment due to resistance or lack of response), and to minimize the impact of anticipatory plan switching.

Individuals are included if they had no previous indication of CML diagnosis or treatment during the washout period, two or more diagnoses consistent with International Classification of Diseases 9th and 10th revision (ICD-9 and ICD-10) diagnosis codes for "Chronic myeloid leukemia, BCR/ABL-positive, not having achieved remission" (205.10 and C92.10, respectively), and at least one branded imatinib order within 12 months of confirmatory diagnosis. Patients meeting the following criteria are excluded: treatment switching within the follow-up period, non-commercial insurance in the pre-Medicare period (Medicaid, VA), subsidized Medicare coverage (dual coverage), and inpatient or outpatient claims listing a diagnosis code consistent with any additional indications for imatinib (31).

We have two key exposures of interest, a "running variable" and a binary "cutoff," or "threshold," variable representing the actual treatment (or what we are interested in estimating the causal effect of) that is a function of the running variable. In this analysis, the running variable is age and the threshold variable is age 65. Elasticity is estimated using the adherence and OOP measures described in Table 1.

Use of secondary data means that IRB approval was not required for this study.

**Empirical strategy**

The estimation of price elasticity of demand (PED) for branded imatinib using observational data is complicated by the presence of multiple sources of bias - primarily, endogeneity associated with plan choice due to individuals most sensitive to price choosing more generous plans either at eligibility or in response to their CML diagnosis. Omitted variable bias is also a concern due to the lack of information on several important confounders in TriNetX data (e.g., socioeconomic status (particularly disposable income) and plan structure). Additionally, selection bias regarding which patients remain on some form of employer-sponsored health plan and which enroll in Part D post eligibility is also present.

To address these concerns, we use a regression discontinuity design (RD) to retrieve the causal effect of Medicare Part D enrollment at age 65 on initial OOP and adherence. If identifying assumptions are met, the exogenous shock of universal Part D eligibility at age 65 should lead to treatment assignment that is "as-if random" (32). These designs also offer the advantage of demonstrating high internal validity in practice and requiring what many consider to be less stringent and more easily testable identifying assumptions than other causal inference methods (33,34).

The core identifying assumption of RD is continuity, which requires that the threshold (age 65 in this case) is itself not endogenous with respect to another "shock" or treatment occurring simultaneously (35). In other words, there is a "smoothness" assumption that, in the absence of universal eligibility for Medicare enrollment at age 65, adherence and OOP would have remained continuous across this time period. Importantly, continuity allows for there to be differences in both observed and unobserved confounders around the cutoff as long as they vary "smoothly" with respect to age (36). We evaluate this assumption using the following variables measured during the baseline period: sex, comorbidities (based on Charlson Comorbidity Index), mean 30-day OOP in the 90 days prior to diagnosis, diagnosis year, and diagnosis month.

Aside from the need for a discontinuous jump in the probability of treatment at the threshold (Medicare Part D eligibility at age 65 in this case), the other key assumption of the RDD is that individuals must not be able to

*precisely* manipulate the threshold for treatment (35). This assumption should hold, as individuals cannot change their age or health status to qualify at any other time.

In a sharp regression discontinuity design (SRD) *all* individuals in the sample above the threshold are eligible for *and* comply with treatment (compliance is 100%). Because individuals may qualify for Part D early at age 62 (through Social Security disability insurance (SSDI) end-stage renal disease, or ALS) and individuals post-65 may choose to not to enroll, this analysis represents a fuzzy regression discontinuity design (FRD). For this design, while there's still a jump in treatment, receipt post-threshold is now probabilistic (greater than 0% and less than 100%) (37).

We implement an FRD using the Wald estimator, which effectively estimates the intention to treat effect (numerator) scaled by the proportion of compliers (denominator) at the threshold. The numerator is estimated using a reduced form RD model that is equivalent to that used for a sharp design. The scaling factor is derived based on the same design, but with the probability of treatment (enrollment in Part D at age 65) as the outcome (35,36,38). The general equation used to estimate each model is provided below.

$$y_i = \gamma * 1(age > 65) + f(age) * 1(age \leq 65) + g(age) * 1(age > 65) + \varepsilon_i(age)$$

where $y_i$ represents a study outcome (OOP, adherence, or Part D enrollment) for patient $i$, $f(age)$ and $g(age)$ are smooth functions of age for individuals below and above the age 65 threshold, respectively, $\varepsilon_i$ represents the error term, and $\gamma$ is the effect of Medicare eligibility at age 65.

The canonical RD relies on the use of a continuous "running" variable, where the presence of a cutoff or threshold splits the sample by treatment status (39). The TDN data, however, do not include age, but rather birth year from which we can derive age using the index date. Because we are estimating adherence over a three-month period, measurement error at age 65 would run the risk of "contamination" in our price elasticity estimates if OOP and adherence were attributed to the wrong treatment.

When the running variable is discrete, as in this case, traditional RD methods become less reliable (40). To address this issue, a "donut RD" (DRD) combined with an "honest confidence" interval approach based on Goldsmith-Pinkham et al and Kolesar et al is adopted (41,42). The DRD design gets around the lack of precision in the running variable by omitting values at the threshold (individuals with an index date sometime in the year they turned 65 in this case) (40).

Bias-adjusted 95% confidence intervals are estimated via "honest" methods, which have guaranteed coverage properties and account for the bias caused by the additional extrapolation required with a discrete running variable and lack of observations at the threshold (42,43). The impact of different scaling parameter values used to estimate the tuning parameter from this method are explored in sensitivity analyses.

Price elasticity of demand is calculated using results from each fuzzy regression discontinuity (FRD) to represent the change in quantity demanded $(Q_{post} - Q_{pre})$ and price $(P_{post} - P_{pre})$ at age 65 among those who enroll in Part D due to expanded eligibility and average adherence $(Q_{pre})$ and OOP $(P_{pre})$ immediately prior to age 65 to estimate the ratio of the % percent change in quantity (90-day adherence) divided by the percent change in price (initial OOP) from the pre- to post-period (see below).

$$\frac{(Q_{post} - Q_{pre})/Q_{pre}}{(P_{post} - P_{pre})/P_{pre}} = \frac{\%\Delta\ adherence}{\%\Delta\ price}$$

We do not consider substitution effects to be necessary since the study period occurs prior to generic imatinib introduction when only costlier, imperfect substitutes (second-generation TKIs) were also available as first-line treatment.

Bootstrap resampling was used to construct standard errors and confidence intervals (44).

**RESULTS**

**Change in Part D enrollment, OOP, and adherence at age 65**

In Figure 1, we estimate the effect of age 65 on Part D enrollment (scaling factor), OOP (reduced form RD), and adherence (reduced form RD). These plots represent *sharp donut regression discontinuity designs* where the exposure is age 65 (i.e., when universal Medicare Part D eligibility takes effect). The results presented are from local linear or polynomial regressions using a 10-year bandwidth on either side of the threshold.

We find that there is a marked and significant increase in Part D enrollment at age 65 of 58 percentage points (95% CI: 0.45 to 0.71). Initial OOP at this age also jumps significantly by $135 (95% CI: $53 - $218). While initial adherence does change as well, the reduction of 4 percentage points (95% CI: -0.12 to 0.05) is small and non-significant.

**Impact of Part D enrollment on OOP and adherence at age 65**

To better understand the effect of Medicare Part D *enrollment* on patients, the previously discussed sharp RD results were used to form the numerator (change in adherence and OOP) and denominator (change in Part D enrollment) of the Wald estimator. Results from this fuzzy regression discontinuity design looking only at individuals that enrolled in Part D due to expanded eligibility at age 65 found a significant increase in average OOP of $232 (95% CI: $102 to $362). Similar to the sharp RD results, the corresponding decrease in average adherence was small, at only 6 percentage points (95% CI: -0.21 to 0.08), and non-significant.

**Price elasticity of demand**

To get an idea of how responsive newly diagnosed CML patients are to the cost for an initial branded imatinib prescription, we combined results from our separate RDs and pre-65 adherence and OOP to estimate the price elasticity of demand (percent change in adherence per 1% change in OOP). We are using OOP here rather than list price because our research question of interest is centered on how patients respond to the price that they actually face when filling their first prescription (i.e., OOP) and how this varies across different benefit structures. The list price for branded imatinib may or may not be changing, but this shouldn't affect our results given that index year is balanced around the threshold (See section 3.4).

For our main results, PED in those individuals that enrolled in Medicare Part D due to expanded eligibility at age 65 were found to be extremely inelastic with respect to price, with a PED of -0.02. These results were non-significant and fairly imprecise (95% CI: -0.056, 0.015), most likely due to our smaller sample size and the uncertainty in our adherence estimate (Table 1).

PED estimates in our ITT population (all individuals eligible for Medicare Part D at age 65)/from our sharp RD were the same as our main results (-0.02). This makes sense intuitively, as the values used in the numerator and denominator for our main results are all scaled by the same factor in the FRD.

**Robustness checks**

In Table 1, we also examine the sensitivity of our results to model specification by estimating discontinuities and PED using different functional forms for parametric and non-parametric models. We show that, while magnitudes are affected, overall model results are robust to polynomial degree and global vs local fit, with estimated PED ranging from -0.004 to -0.021. Only the estimate for OOP in the local cubic model moved from highly significant to non-significant. This is not surprising given polynomial fits over the $2^{nd}$ degree have been demonstrated to perform poorly at the boundaries, tend to overfit the data, and are not recommended in practice (37).

The main assumption that we make when calculating Honest CIs relates to the value that we assign to our "tuning parameter," which limits the extent to which the function that we're using to extrapolate from our boundary point at age 64 and 66 to age 65 can change (42,43). We follow the approach described by Kolesár et al and implemented by Goldsmith-Pinkham et al in estimating this parameter where, because we want to account for how wrong we could potentially get our extrapolation, we use the coefficient from a quadratic global regression model and multiply it by a pre-determined scaling factor (four in this case) (41,42). The results of varying this scaling factor for both the adherence and OOP analyses are presented in Table 1 and indicate that our estimates are robust with respect to choice of scaling factor; although, the change in OOP was no longer significant at the 0.01 level (but still significant at the 0.05 level).

In Figure 2, we test the core identifying assumption of the regression discontinuity design – that there is "smoothness" in other observed characteristics at age 65. If this assumption is violated, it indicates that the discontinuities present in the data may be due to other factors that jump discontinuously at age 65 other than enrollment in Medicare Part D. Based on figure 2, there do not appear to be any discontinuities in other covariates at the threshold and observed effects are non-significant.

While can empirically test observed characteristics, we have to assume that there are no unobserved "shocks" or treatments that would cause either OOP or adherence to jump at the age 65 threshold. An important potential shock to consider is the impact of retirement, as this traditionally occurs at age 65 in the US. If retirement is the primary factor leading to decreased income and subsequent changes in other lifestyle factors, this could also drive any observed discontinuities in OOP and subsequently adherence. Previous work and current Bureau of Labor Statistics data, however, indicate that retirement age in the US has become more of a smooth function over time (41,51).

We further test the robustness of analysis results by varying the bandwidth used in the local linear regressions for OOP and adherence. Figure A1 demonstrates that OOP results were insensitive to bandwidths ranging from 5 to 15, with significance and directionality consistent across all values. Adherence results are robust with respect to significance (none of the specifications resulted in a significant difference). While the directionality of the effect did change from slightly negative to slightly positive for bandwidths of 5 and 15, this is likely a reflection of the smaller sample available for the former leading to increased variance/extremely wide 95% CIs and increased bias due to less comparable treatment and comparison groups in the latter.

Placebo tests are additionally performed to assess whether the discontinuity in outcomes observed at age 65 could potentially be due to random noise in the data. Figure A2 indicates that results are robust to these placebo thresholds, with the main finding of a significant discontinuity in average OOP holding only at age 65. For adherence, the null hypothesis of no effect size could not be rejected for any of the placebo thresholds. More broadly, the largest effect sizes for both outcomes were seen at the age 65 threshold.

Figure A3 provides average outcomes and enrollment by age from age 50 to age 80 to give a better sense of overall trends in the data using global polynomial (quadratic) regressions.

**CONCLUSION**

For policy makers and payers to strike the optimal balance between cost sharing and financial risk protection, it's essential that robust estimates of the downstream impact of high drug prices on patients are available. This is particularly true in the Medicare Part D program, which has seen an explosion in spending in recent years due to the increase in high-priced specialty drugs entering the market (2). The coinsurance-based benefit design in Part D means that patients bear a significant cost for these medications as well due a lack of any spending cap in the catastrophic coverage phase (3,4). In addition to the financial burden placed on patients, these high OOPs can have serious consequences in terms of non-initiation and sub-optimal adherence. This has important implications for oral cancer therapies like imatinib, where adherence as high 90% is required to achieve and maintain remission.

In this analysis, we implement a fuzzy donut regression discontinuity design to assess the impact of moving onto Medicare Part D at age 65 on average initial OOP and adherence for branded imatinib in newly diagnosed CML patients. Combining results from each, we estimate how much OOP increases affect patient adherence. Our finding that there is a large and significant increase in both Part D enrollment (58 percentage points) and average OOP ($232) is in line to what has been found in similar studies (10,12,45,46). Interestingly, although associational studies have identified a link between OOP and adherence, we find only a small and non-significant reduction of 6 percentage points at age 65 (10,22,26–28). This difference is potentially due to our ability to account for the endogeneity associated with plan choice, the distortionary effect of coupon use, and our new user design. Results from both analyses were robust to placebo cutoffs and different model specifications.

The corresponding PED of only -0.02 suggests that CML patients are extremely inelastic with respect to their initial branded imatinib prescription cost. This is largely consistent with the broader literature on the price elasticity of demand for specialty drugs (5,6,29,47). This PED value is much lower than the estimates from Jung et al for leukemia patients (-0.96 to -0.99), but this is likely due to differences in study scope. Given that Jung et al is looking at Part D patients at all ages, for all types of leukemia, and includes both prevalent cases and non-initiators, it is not unexpected that results would be different. This would be particularly true if the conditional expectation function for each outcome given age is highly heterogenous since Jung et all include all patients over 65 (17).

While there are significant advantages to implementing an RDD, external validity is limited by its inherently local nature. This means that results from this analysis are relevant only to patients who enroll in Medicare Part D at age 65. Because we are estimating OOP and PDC longitudinally, there is a chance we may capture some plan switching in response to CML diagnoses/ branded imatinib. The short length of follow-up (90 days) and restricted window for plan switching (October 15 to December 7 each year) should limit the impact of any anticipatory behavioral changes.

Our results suggesting that high OOP is not a determinant of nonadherence in the approximately 20-30% of patients that were observed to exhibit this behavior indicate that there may be other factors driving noncompliance in this population. The non-responsiveness of CML patients to initial OOP suggests that high-cost sharing plans, particularly those implementing price-based coinsurance, may be putting patients at unnecessary financial risk. Results highlight the importance of future work in focusing on other potential causes of nonadherence in the older CML population and assessing the impact of high-cost-sharing benefit designs on patient costs and adherence for other specialty drugs.

# Tables and figures

## Table 1. Key study variables

| Variable | Estimation method |
|---|---|
| **_Exposure_** | |
| Age | Because TDN does not include age at claim information, year of birth will be used to get an approximate estimate of age at baseline (year of index date – birth year). |
| **_Outcomes (standardized to 90-day follow-up period)_** | |
| Initial adherence | Based on proportion of days covered (PDC) - defined as the number of days supplied of branded imatinib over the follow-up period, with maximum adherence capped at 100%. |
| Initial out-of-pocket payment (OOP) | OOP includes total patient responsibility for the prescription minus coupon value (where applicable). A 90-day fill uses just the OOP for that initial fill, while the OOP for multiple 30-day fills is averaged and standardized to match the 3-month follow-up time. |

**Table 2. Primary results and sensitivity to model specifications**

| Model | Main Analysis (n = 1,416) | Global/parametric | | | Local | | Honest CI Scaling Factor | |
|---|---|---|---|---|---|---|---|---|
| | | Linear | Quadratic | Cubic | Quadratic | Cubic | 2 | 6 |
| First Stage | 0.582*** (0.067) | 0.650*** (0.042) | 0.555*** (0.066) | 0.523*** (0.097) | 0.582*** (0.067) | 0.505** (0.199) | 0.582*** (0.054) | 0.582*** (0.078) |
| Fuzzy – OOP | $232*** ($66) | $141.53*** ($48.17) | $288.152*** ($76.999) | $247.370** ($105.945) | $315.785** ($123.493) | $307.010 ($257.940) | $232*** ($48) | $232** ($91) |
| Fuzzy – Adherence | -0.063 (0.075) | -0.030 (0.042) | -0.045 (0.074) | -0.068 (0.110) | -0.063 (0.075) | -0.015 (0.254) | -0.063 (0.059) | -0.063 (0.094) |
| PED | -0.020 (-0.056, 0.015) | -0.016 (-0.038, 0.021) | -0.012 (-0.029, 0.011) | -0.021 (-0.049, 0.007) | -0.009 (-0.031, 0.063) | -0.004 (-0.083, 0.264) | -0.020 (-0.042, 0.017) | -0.020 (-0.077, 0.008) |

*** = significant at the 1% level; ** = significant at the 5% level. Standard errors in parentheses (bootstrap confidence intervals for PED). All estimates rounded to three decimal places.

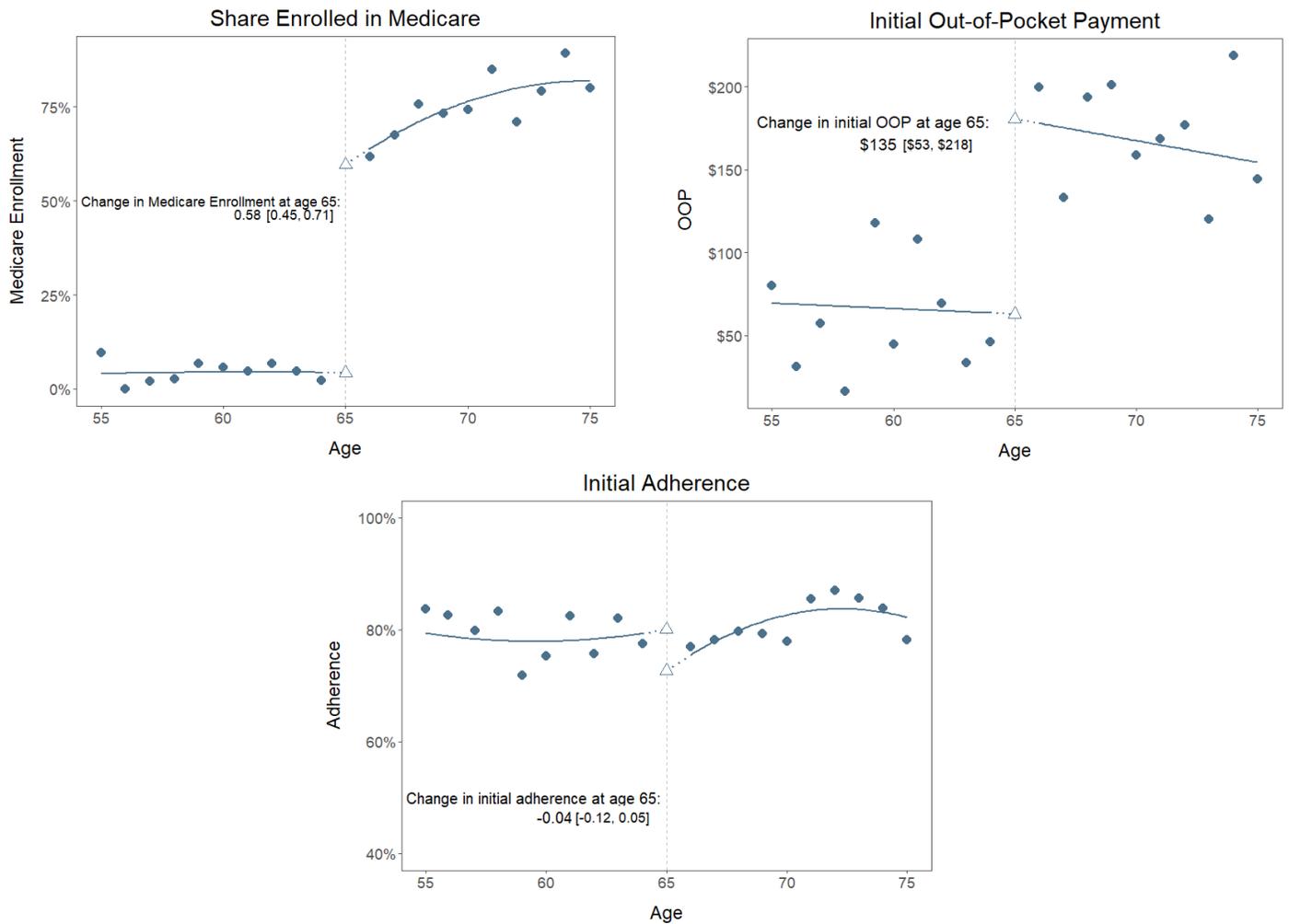

**Figure 1: Effect of age 65 on Part D enrollment, OOP, and adherence**

Each circle represents the average outcome by age at index date from the data. The hollow triangles at age 65 are estimates of values at age 65 from local regressions as the threshold is approached from above and below. The lines are fitted values from a local linear regression for OOP and local polynomial regressions (quadratic) for Part D coverage and adherence with a triangular kernel and a bandwidth of 10. The dotted lines represent the additional extrapolation required for a donut RD due to dropping observations at the threshold (age 65). Adherence is based on proportion of days covered. Each plot reports the sharp RD effect with bias-adjusted 95% confidence intervals.

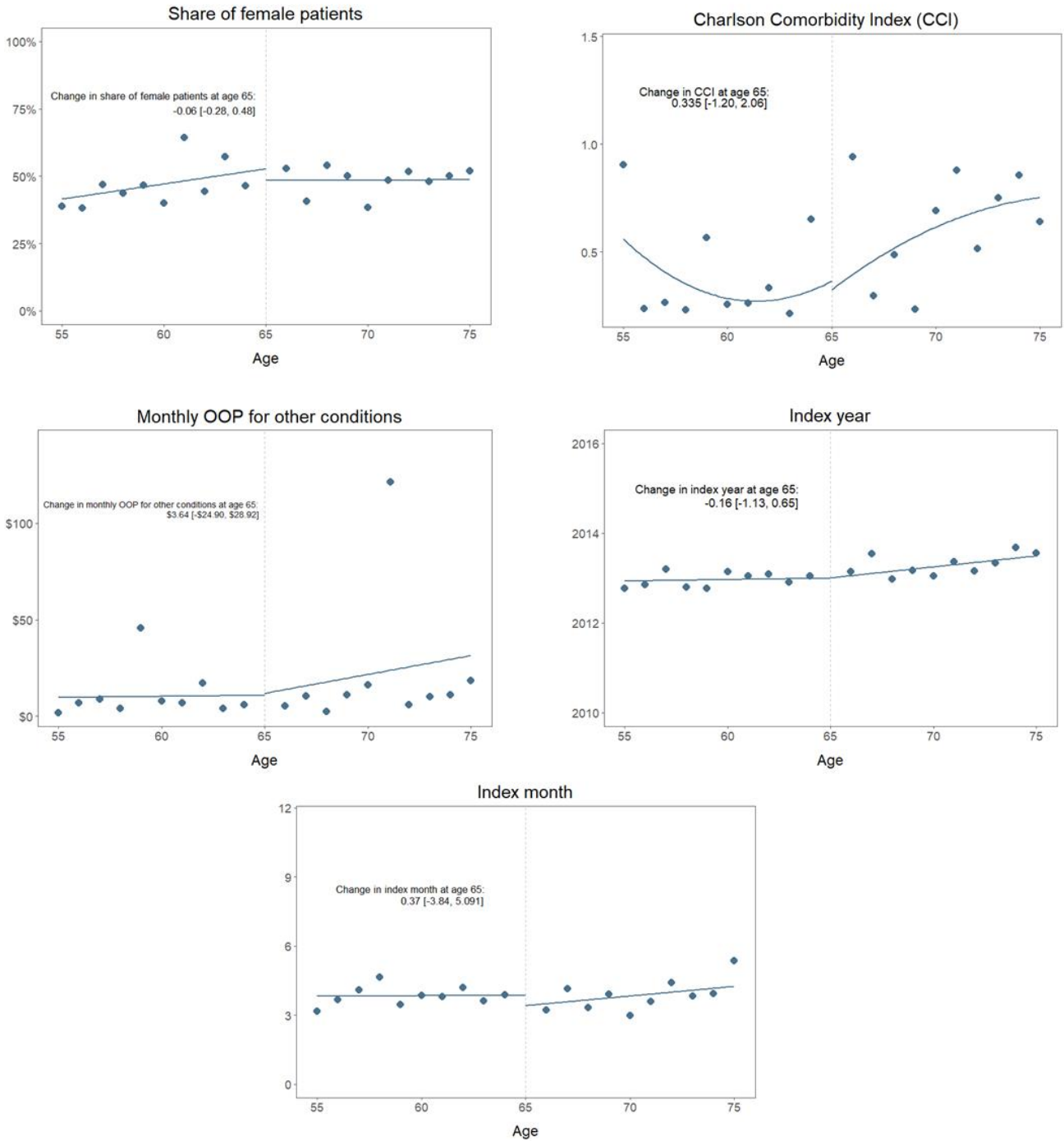

**Figure 2: Regression discontinuity plots and estimates for pre-treatment variables at age 65**

Each circle represents the average outcome by age at index date from the data. The lines are fitted values from local linear or polynomial regressions with triangular kernel and bandwidth of 10. Each plot reports the sharp RD effect with bias-adjusted 95% confidence intervals.

**Supplementary materials**

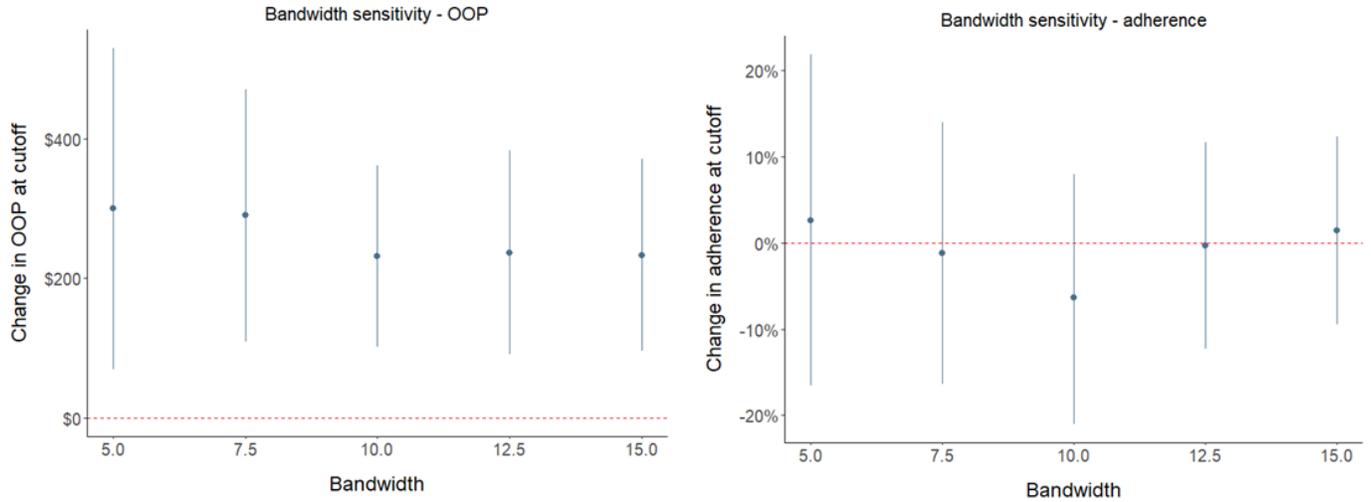

**Figure *A*1: Sensitivity of results to bandwidth selection**

These plots represent the impact of various bandwidths (selected based on Kolesár et al *(42)*), with circles representing point estimates of average effects, vertical lines representing the corresponding 95% confidence intervals, and dashed red line representing a null finding of no effect.

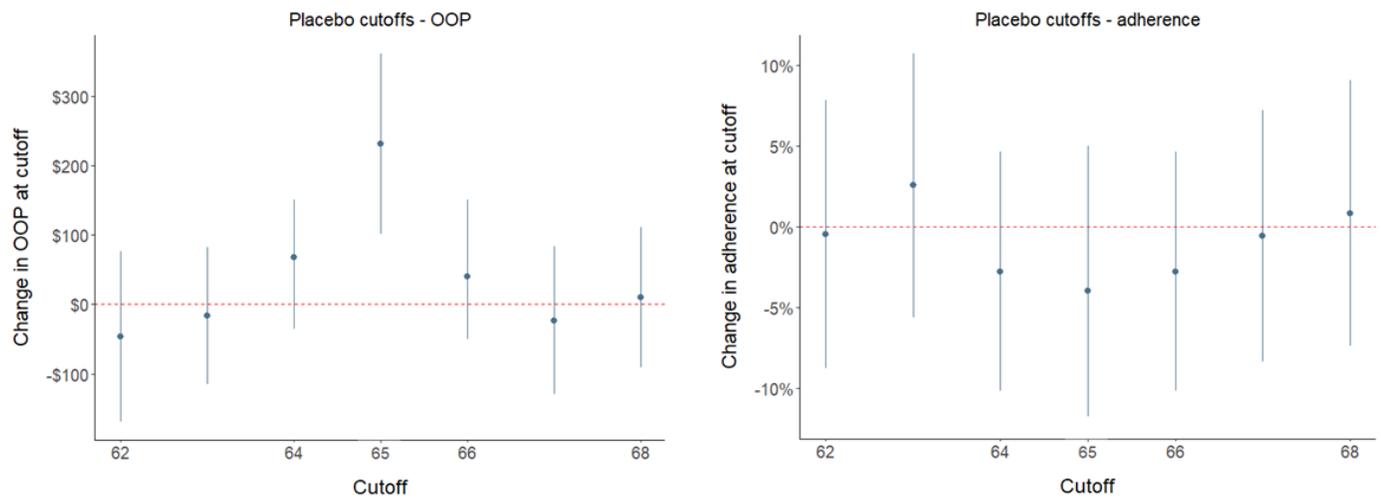

**Figure A2: Robustness of results to placebo thresholds**

These plots represent the impact of using an RD threshold other than age 65 on results, with circles representing point estimates of average effects, vertical lines representing the corresponding 95% confidence intervals, and dashed red line representing a null finding of no effect.

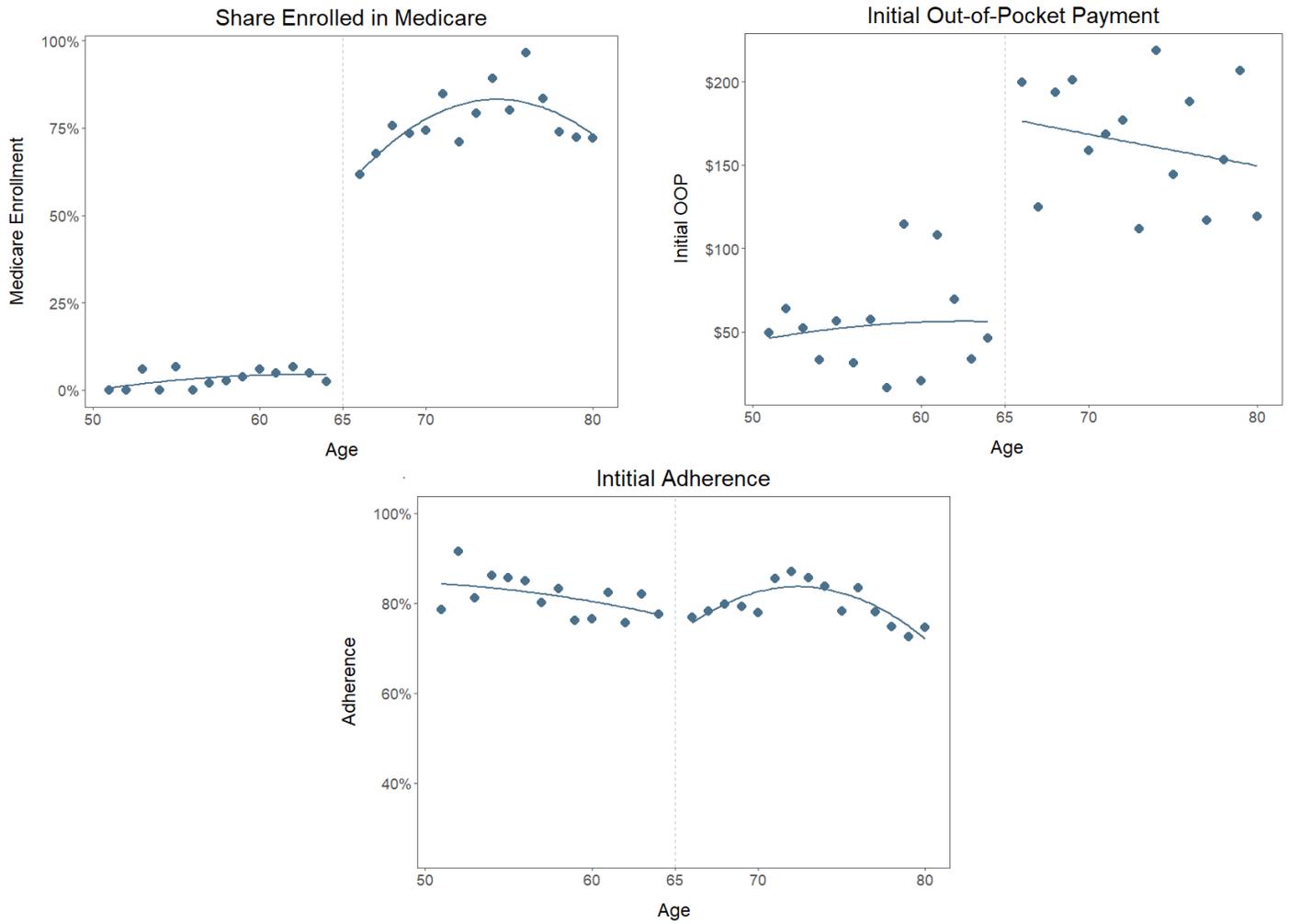

**Figure A3: Global trends in Part D coverage, OOP, and adherence**

Each circle represents the average outcome by age at index date from the data. The lines are fitted values from a global polynomial (quadratic) regression.